# An Exploratory Study of Mobile Computing Use by Knowledge Workers

Paul Prekop
*DSTO Fern Hill, Department of Defence, Canberra ACT 2600*
*paul.prekop@dsto.defence.gov.au*

## Abstract

*This paper describes some preliminary results from a 20-week study on the use of Compaq iPAQ Personal Digital Assistants (PDAs) by 10 senior developers, analysts, technical managers, and senior organisational managers. The goal of the study was to identify what applications were used, how and where they were used, the problems and issues that arose, and how use of the iPAQs changed over the study period. The paper highlights some interesting uses of the iPAQs, and identifies some of the characteristics of successful mobile applications.*

## 1. Introduction

Personal Digital Assistants (PDA) are a current example of highly mobile hand-held computing. Understanding how and where these devices are used, and what characterises a usable mobile application, is vital for practitioners and researchers developing mobile applications (Landay & Kaufmann, 1993).

This paper describes some preliminary results from a 20-week study on the use of Compaq iPAQ PDAs by 10 developers, analysts, technical managers, and senior organisational managers. The goal of the study was to identify what applications were used, how and where they were used, the problems and issues that arose, and how use of the iPAQs changed over the study period.

This paper describes some of the ways the PDAs were used by the study participants, and identify some of the characteristics of successful mobile applications.

## 2. Related Work

Numerous studies have explored the use of highly mobile devices, like hand-held computers and PDAs, to support specific applications. For example, Waycott & Kukulska-Hulme (2003), studied the use of PDAs for reading course material. Newcomb, Pashley & Stasko (2003), studied the use of PDAs to support grocery shopping, and Spain, Phipps, Rogers & Chaparro (2001) explored the utility of PDAs as a data collecting device.

The study described here is different; it focuses on the overall utility of PDAs to a particular class of users, rather than focusing on the utility of the device for a particular application.

## 3. The Study

The study consisted of 10 participants, and ran for 20 weeks. All participants were well educated, (all had bachelors degrees, some had Masters or PhDs), highly IT literate (self-described), senior developers, analysts, technical managers, and senior organisational managers. Some had previous experience with mobile computing.

Participants received a new Compaq iPAQ (5 participants received the 3700 and 5 received the 3800 model) for their exclusive use. They received no instructions or training on how to use the device, and no additional software for the device, apart from the software supplied with the iPAQ by Compaq. All had access to professional system administrators able to support the iPAQs. All participants successfully synchronised the iPAQ with their main computer.

Since we did not fund the purchase of the iPAQs used in the study, we were unable to control who was selected for the study. However, the participants selected reflected a good cross section of senior staff and managers within the organisation.

**Table 1. Study timeline**

| Week | Data Collected |
|---|---|
| 1 | Initial Interviews, Demographic & Work Baseline Survey |
| 3-4 | Christmas/New Year Holiday Period |
| 7 | Trend Survey 1 |
| 10 | Trend Survey 2 & Mid-Point Interviews |
| 13 | Trend Survey 3 |
| 16 | Trend Survey 4 |
| 19 | Trend Survey 5 & End-Point Interviews |
| 20 | End-Point Interviews |

The participants were surveyed and interviewed throughout the 20-week study period (see Table 1). The 5 Trend Surveys measured how frequently the iPAQs were used during the previous 3-week period (except for Trend Survey 1, which also measured the holiday period), what applications were used, where the device was used, and how the device was customised.

The interviews were semi-structured and lasted about an hour each. They were used to gain a richer insight into the participants' experiences using the iPAQ.

The study methodology resulted in a very large and rich set of data. However, the extremely small sample size and short duration of the study, means the results can only be considered exploratory and qualitative, and may not be generalisable outside of the sample.

## 4. Results and Discussion

The Trend Surveys asked participants to record how frequently they used the device over the previous 3 weeks. Device use was measured on a 5-point end-anchored scale; the scale was anchored at *never used (1)* and *very frequently used (5)*. Figure 1 shows the mean of the device use scale over the study period.

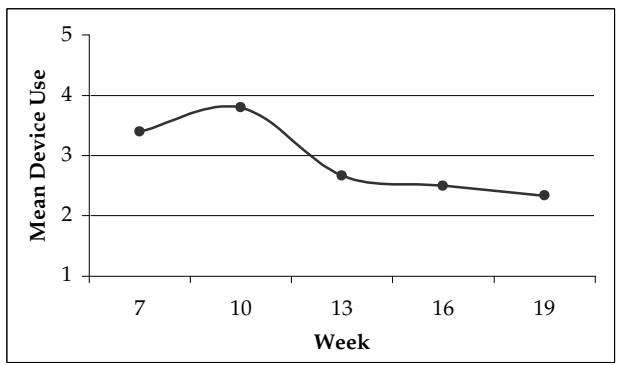

**Figure 1. Device use over study period**

As shown by Figure 1, the mean use of the device peaked during the first 7 to 10 weeks of the study (Trend Surveys 1 and 2), before dropping for the remainder of the study. Data collected on the applications used follows this trend, with application use peaking during the first 7 to 10 weeks (Trend Surveys 1 and 2), before settling to a sustained level, or dropping off all together. Data collected on device customisation and the addition of software, reflects this trend again, with most customisation taking place during the first 7 to 10 weeks (Trend Surveys 1 and 2).

The mid-point and end-point interviews revealed that during the first 7 to 10 weeks, the participants spent a considerable about of time configuring and re-configuring the device, installing software on it, and most importantly, learning about the device's capabilities and features through use. By the mid-point interviews (week 10), most participants felt they knew enough about the device to be able to use it to support their work. Also, by this point, several participants had started to critically evaluate the device's usefulness to their work.

The Trend Surveys also asked participants to list (as a percentage) how frequently they used their new iPAQ, their main work computer, or other methods/technologies to support various activities. Table 2 shows the mean percentage of time the iPAQ was used to support various activities at week 7 (Trend Survey 1) and week 19 (Trend Survey 5)[1].

**Table 2. PDA use for various applications**

| Applications | WK 7 (TS1) | WK 19 (TS5) |
|---|---|---|
| Calendar/Diary | 29.6% | 22% |
| Note taking in meetings | 26.5% | 2% |
| Note taking while thinking | 8% | 6% |
| Contact information | 16% | 17.5% |
| E-Mail | 4.1% | 4% |
| To-do Lists | 20% | 11% |
| Other Reading | 9.5% | 0.5% |
| Gathering Information | 4.1% | 0% |
| Writing | 3.1% | 2% |
| Other | 7.5% | 3% |

The comparison between weeks 7 and 19 shows which applications the participants started using (week 7), then after 12 weeks of experience with the device, what applications they still found the device useful to support (week 19). As shown by Table 2, some applications were used continuously throughout the study period, while some applications went from high use at the start of the study, to little or no use by the end of the study. The mid-point and end-point interviews were used to gain an understanding of why particular applications were used, while the use of other applications dropped off all together.

### 4.1. Note Taking in Meetings

The use of the iPAQ to take notes in meetings was the one application that changed the most significantly from week 7 to week 19.

The initial interviews revealed that almost all the participants felt the iPAQ would be useful for note taking in meetings, enabling them to take better meeting notes, and making it easier to translate notes to text or Word files, and to manipulate them on a computer. Initially almost all the participants used the device to take notes in meetings. However by week 19, almost none of the participants used the device in this way.

Most participants who attempted to use the device to take notes in meetings used the iPAQ's handwriting recognition, either free text or character by character.

The interviews revealed that a common difficulty the participants had using the iPAQ to record meeting notes was the amount of attention and concentration it took to input information using handwriting. As a result, the participants found they were concentrating more on entering information into the iPAQ than on participating in the meeting.

---

[1] The sample size was too small for meaningful significance tests to be performed.

Note taking is an important activity for knowledge workers. As discussed by Kidd (1994), the process of note taking is often far more important than the resulting notes. Note taking seems to in some way fix ideas or information into the knowledge worker's mind.

Only one participant really found the iPAQ useful for note taking during meetings. However, his use of the device was very different to most other participants. He would often use a small external keyboard to type his notes during meetings, rather than use handwriting. Also, before receiving the iPAQ, he had a pre-existing process for typing his handwritten meeting notes into his main computer, and a system for managing the electronic version of his meeting notes.

While none of the participants was completely satisfied with the iPAQ's handwriting interface, it didn't prevent them from entering information into the iPAQ for other activities (the frequency of use of the iPAQ for note taking while thinking or reading, and writing activities changed little from week 7 to week 19). However, use of the handwriting interface while also participating in a meeting seemed to be too demanding.

### 4.2. Opportunistic E-mail and Reading

While not a high-use application, e-mail was constantly used over the study period, and e-mail use highlights an important characteristic of mobile computing.

None of the participants had a wireless network connection on their iPAQ, and the only e-mail they could access was a disconnected snapshot copied from their main computer. Because of the limited memory available on the iPAQ, the e-mail snapshot was generally limited to 4 to 6 weeks.

The participants' use of e-mail was mostly opportunistic, and used during dead time; time spent travelling or waiting. They would use the time to catch up on reading background, or low importance e-mails (including e-mail lists) they hadn't read while in the office. Few participants attempted to write or reply to e-mail on the iPAQ, and the few that did tended to reply with very simple, short answers. Any e-mail requiring a detailed response, or in-depth reading or consideration, was flagged and dealt with when the participant was at a more suitable location (office, home, hotel) and/or when they had access to a desktop or laptop.

As well as e-mail, many participants made use of the AvantGo (www.avantgo.com) news service (shipped with the iPAQ by Compaq). AvantGo provides news feeds from general and technology focused news services, specifically formatted for the iPAQ. None of the participants who made use of AvantGo obtained the information provided by AvantGo via any other means. As with e-mail, most of the AvantGo reading was opportunistic, and done during dead time while mobile.

Both e-mail use and AvantGo reading highlight an interesting characteristic of opportunistic use of the device. The use of the device was *cognitively light* – the participants were unable or unwilling to invest significant mental effort to use the application or information. Instead they used it to catch-up on background or low importance information, and deferred dealing with any information that required in-depth reading or consideration until a more suitable time and/or place.

An interesting question is: was this behaviour due completely to the location of use, i.e. was the device used in locations where in-depth reading or consideration is not practical, or was this behaviour due to the iPAQ's form factor, i.e. its small screen and handwriting interface? The study didn't collect any data that could definitively address this question.

Another interesting characteristic of the iPAQ that affected its use while mobile was the set-up time needed to be able to use the device to access e-mail or to read. Many of the participants who had an iPAQ also had a laptop, and many would take both their laptop and iPAQ while mobile. While the iPAQ was used opportunistically to access e-mail or read, the laptop, which generally contained the same information, wasn't. The interviews revealed that the laptop's form factor, and its much longer set-up time (the time needed to unpack it, the time needed for the laptop to start-up, the time needed to get to the required application, and the right place in the application), meant that most participants didn't feel it was worth using the laptop in the same way.

### 4.3. Mobilising Information

A frequently used source of information that the iPAQs mobilised (i.e. made available while mobile) was calendar/diary information. As shown by Table 2 around 20-25% of the participants total calendar/diary use was via their iPAQ, and calendar/diary use was sustained over the study period.

Adding mobility to calendar/diary information required very little effort for the participants. Almost all of them used Microsoft's Outlook (a combined, e-mail, calendar/diary, contact information, and to-do list application) to manage their calendar on their main work computer. Outlook is also available on the iPAQ. Whenever the participant's iPAQ was connected to their main computer, the calendar/diary information contained on the two devices was synchronised.

While all the participants felt that mobilising calendar/diary information was useful, and added some value, few felt it added greatly to their overall productivity, and none of the participants described it as a "must have" application. Before the introduction of the iPAQ, participants who needed a mobile form of their calendar/diary simply printed the week's or month's calendar from Outlook.

As well as mobilising calendar/diary information, Outlook also mobilised contact information – names, phone numbers, addresses and e-mail addresses. While not as frequently used as the calendar/diary information, the impact of mobilising contact information seemed to be significant for several participants. They generally didn't have this information in a mobile form before the introduction of the iPAQ and they were often unable to predict what contact information would be useful while mobile; hence they found it difficult to make the contact information mobile via other means, for example printing it.

An interesting use of the iPAQ to mobilise information was its use as an *information carryall*, a simple, lightweight and convenient way of carrying diverse information for specific purposes. Examples of this included: specifically moving background or supporting documents (generally Word or Power Point) to the iPAQ in preparation for an activity; using the iPAQ to hold background e-mails relevant to an activity; moving all the information needed for an extended trip – flight details, hotel details, meeting agenda, and so on, on to the iPAQ. Use of the iPAQ in this way was infrequent, and tended to vary with the participants work schedules and needs. The information held wasn't generally added to or processed in any way; it was simply read. This kind of reading is different from the reading described in the previous section. The issues associated with opportunistic reading – unwillingness to invest mental effort or significant time – didn't seem to apply to this kind of reading.

For the participants who used the iPAQ in this way, it generally removed the need to carry paper copies of the needed documents. Most of the participants who used the device in this way, simply used the native iPAQ file structure and the standard file viewers to manage and view the information, or simply relied on the e-mail application.

A key issue in mobilising information, is what information adds impact while mobile? This section has described three different classes of information, all with varying levels of utility to mobile users. Mobilising contact information seemed to have the most significant impact because contact information was innately useful while mobile, and the participants found it difficult to mobilise without the iPAQ.

## 5. Limitations and Conclusions

This paper has described some of the ways knowledge workers used highly mobile PDAs, and highlighted some problems with the current generation of PDAs.

One problem highlighted was using the iPAQ to take notes in a meeting. It seems that the handwriting interface takes considerable mental effort to use, and in a meeting the participants found their attention was on the iPAQ and not on participating in the meeting.

Opportunistic use of the iPAQ was also described. When used opportunistically, the study shows that the device or application can't demand in-depth reading, consideration or attention by users, or long set-up times.

Finally, this paper described examples of mobilising information. Highly useful, mobilised information seems to be information that has an innate utility while mobile, and that can't easily be mobilised in other ways.

The study described was limited by its extremely small sample size and short duration, meaning the results described can only be considered exploratory and qualitative, and may not be generalisable outside of the sample.

Also, none of the iPAQs used in the study had a wireless network connection. It is likely that wireless networking would have had an impact on the way the participants used the device, especially the use of e-mail, reading, and using the device as an information carryall.